\documentclass[fleqn,usenatbib]{mnras}

\usepackage{graphicx}

\title[Origin of Stellar Flares on M dwarfs]{Probing the Origin of Stellar Flares on M dwarfs Using TESS Data Sectors 1 - 3}

\author[L. Doyle et al.]{
L. Doyle,$^{1,2}$\thanks{E-mail: lauren.doyle@armagh.ac.uk}
G. Ramsay,$^{1}$
J. G. Doyle,$^{1}$ 
K. Wu$^{3}$
\\
$^{1}$Armagh Observatory and Planetarium, College Hill, Armagh, BT61 9DG\\
$^{2}$Mathematics, Physics and Electrical Engineering, Northumbria University, Newcastle upon Tyne, NE1 8ST\\
$^{3}$Mullard Space Science Laboratory, University College London, 
 Holmbury St Mary, Surrey RH5 6NT
}

\date{Accepted 2019 August 06. Received 2019 July 29; in original form 2019 May 29}

\pubyear{2019}

\begin{document}
\label{firstpage}
\pagerange{\pageref{firstpage}--\pageref{lastpage}}

\maketitle

% Abstract of the paper
\begin{abstract}
Detailed studies of the Sun have shown that sunspots and solar flares are closely correlated. Photometric data from Kepler/K2 has allowed similar studies to be carried out on other stars. Here, we utilise TESS photometric 2-min cadence of 167 low mass stars from Sectors 1 - 3 to investigate the relationship between starspots and stellar flares. From our sample, 90 percent show clear rotational modulation likely due to the presence of a large, dominant starspot and we use this to determine a rotational period for each star. Additionally, each low mass star shows one or more flares in its lightcurve and using Gaia DR2 parallaxes and SkyMapper magnitudes we can estimate the energy of the flares in the TESS band-pass. Overall, we have 1834 flares from the 167 low mass stars with energies from $6.0\times 10^{29}$ - $2.4\times 10^{35}$~erg. We find none of the stars in our sample show any preference for rotational phase suggesting the lack of a correlation between the large, dominant star spot and flare number. We discuss this finding in greater detail and present further scenarios to account for the origin of flares on these low mass stars. 
\end{abstract}

% Select between one and six entries from the list of approved keywords.
% Don't make up new ones.

\section{Introduction}

For over nine years, the Kepler mission provided a wealth of time variability information for several hundreds of thousands of stars which provided a wide range of advances in stellar astrophysics and exoplanet research \citep{Borucki2010}. With the loss of its second reaction wheel in 2014, Kepler was re-purposed as K2 and began to take observations of fields along the ecliptic for $\sim$70 - 80 days. However, in October 2018, it ran out of consumables and NASA announced the retirement of the satellite, ending its mission.  

Kepler's successor, the Transiting Exoplanet Survey Satellite, (TESS), was launched in April 2018, with a core mission to obtain high quality, month long, lightcurves of stars brighter than $\sim$13 mag. Unlike the Kepler/K2 missions it will cover nearly the whole sky over the course of its two year prime mission \citep[see,][for details]{Ricker2015}. It will target 500,000 stars, with a focus on nearby G, K and M type with approximately 1,000 of the closest red dwarfs being included in the 2 min cadence programme. 

The lightcurves of low mass M dwarfs can show periodic changes in their brightness as the star rotates. This is widely thought to be the result of a dominant, large starspot which is cooler than its surroundings rotating in and out of view \citep{mcquillan2013measuring}. From observations of the Sun, we know flares typically originate in active regions which host spots so, it is natural to  expect flares to originate also from the prominent starspots in low mass M dwarfs . However, recent studies, such as \cite{ramsay2013short}, \cite{davenport2014kepler} and \cite{doyle2018investigating}, have shown evidence seriously challenging this view. There was no correlation found between the flare number and the rotational phase in any of the M dwarfs observed using Kepler or K2, suggesting flares could occur without the association with a large, dominant starspot.

In our previous study, \cite{doyle2018investigating}, (henceforth Paper I), we used K2 short cadence (1 min) data to investigate the rotational phase of flares in a sample of 34 M dwarfs. Utilising a simple chi-squared test we investigated whether the phase distribution of the flares was random and concluded none of the stars in the sample show any preference for rotational phase. Our result suggests flares on low mass M dwarfs maybe generated through a different mechanism than present in our Sun. 

In this paper we use TESS short cadence (2 minute) photometric data from a selection of M dwarfs made in Sectors 1 - 3 with a prime goal of investigating whether there is a preferential rotational phase for flares from these low mass stars. This sample will be compared to our previous K2 study, were we will discuss in greater detail the potential causes of our findings.

\section{M dwarf Sample Selection}

There are a number of strategies for identifying active low mass stars in TESS data. For instance \citet{Gunther2019} searched for flares in all of the 2 min cadence TESS data and then used the temperature extracted from the TESS Input Catalog \citep{stassun2018tess} to identify cool main sequence stars. Here, we have identified stars which have been observed in 2 min cadence mode using TESS and have a MV spectral type in the {\tt SIMBAD} catalogue \footnote{\url{http://simbad.u-strasbg.fr/simbad/}}. 
By considering the original publications which provided the spectral classifications in the {\tt SIMBAD} catalog, we are confident the stars in our sample have a spectral type accurate to within one spectral subclass. We also used the TESS Input Catalog to remove stars which were likely giants and wrongly classed in the {\tt SIMBAD} catalogue using the luminosity and radii values. 

Our target stars were also cross-referenced with the SkyMapper Southern Sky Survey \citep{wolf2018skymapper}. Those which did not possess Gaia DR2 \citep{gaia18} parallaxes or SkyMapper data were not considered further. The SkyMapper multi-colour magnitudes were converted to flux and then fitted using a polynomial producing a template spectrum which was convolved with the TESS band-pass to derive the stars quiescent flux. The Gaia parallaxes were inverted to provide distances to each star which was used to determine the quiescent stellar luminosity.  

\begin{figure}
    \centering
    \includegraphics[width = 0.47\textwidth]{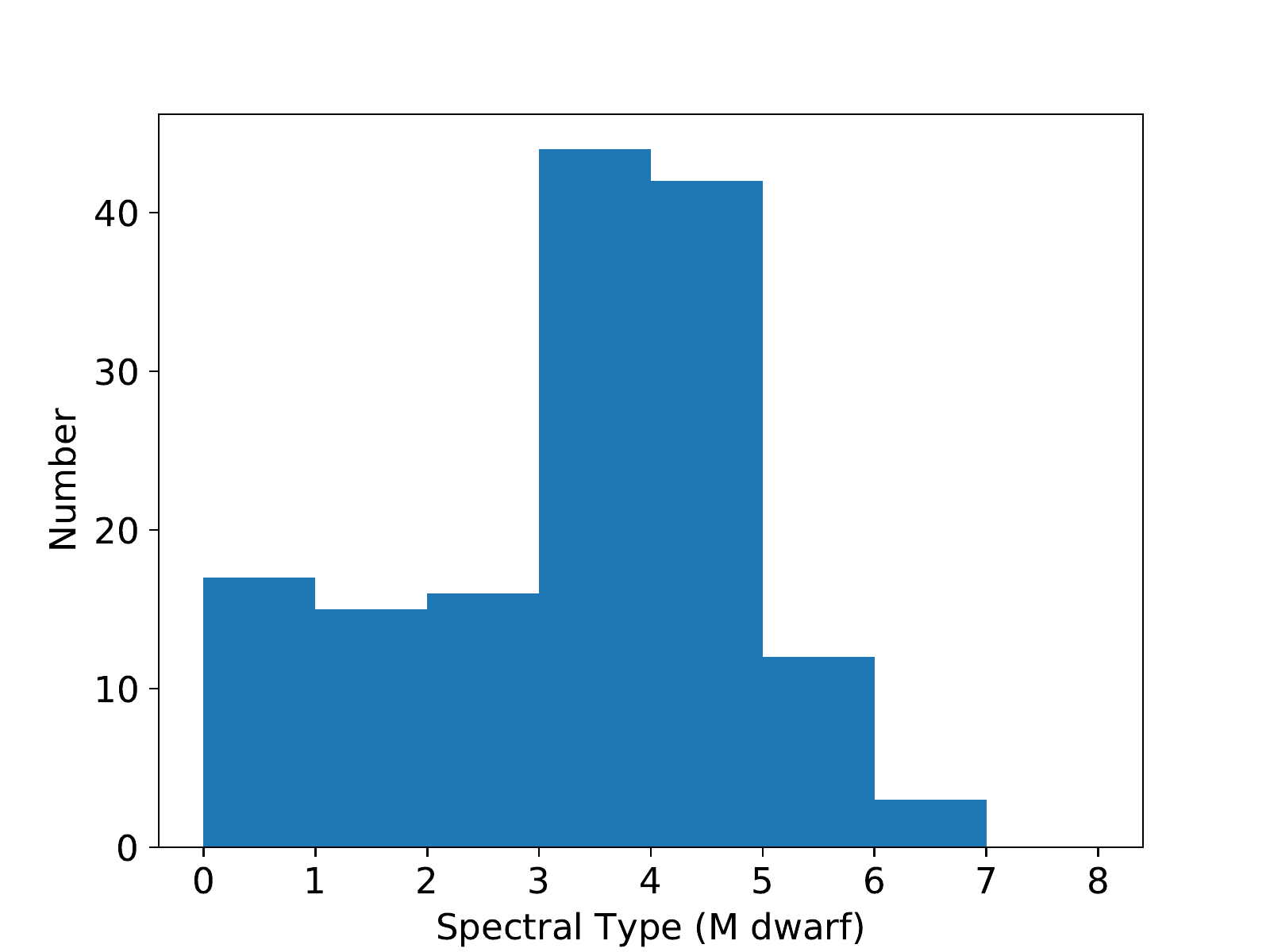}
    \caption{A histogram showing the spread of M dwarf spectral types within our TESS 2-min cadence sample.}
    \label{spect_hist}
\end{figure}

The final sample of low mass stars observed at 2-min cadence with TESS consists of 167 M dwarfs. Each Sector is observed for $\sim$ 27 days with 28 percent of the sample being observed in more than one sector. The complete list of our low mass star TESS sample, including a range of stellar properties, is provided in Table \ref{stellar_properties} and Figure \ref{spect_hist} shows the spread of the spectral types within our sample. 

\section{TESS vs. Kepler/K2}

There are various factors which we need to take into account in our analysis. First, the TESS camera's have a 10.5 cm aperture compared to Kepler's 95 cm diameter mirror. For the same magnitude, TESS will therefore provide lightcurves with an rms noise several dozens of times greater than Kepler. The CCD's in the Kepler detector had a pixel scale of 4$^{''}$ -- this contrasts with 21$^{''}$ in TESS. Indeed, the 90 percent encirclement radius for a stars flux in TESS is 42$^{''}$ \citep{Ricker2015}, which compares with a 95 percent encirclement radius in Kepler of 8.4$^{''}$. There is therefore a potential issue of dilution of the flux of the low mass star with other stars falling into the same aperture.

We therefore used the Gaia DR2 \citep{gaia18} to search for stars within 42$^{''}$ of the stars in our initial target list. All those targets which had another star with a magnitude of up to 1.5 mag fainter (1/4 in flux) were flagged. Of the 167 stars in our list, 46 had a star which was up to 1.5 mag fainter and within 42$^{''}$ of the target. Of these, nine were actually brighter than the target. Whilst this does not affect our prime goal of investigating the rotational phase of the flare it makes us less sensitive to lower energy flares than for stars with no spatially coincident stars and we will underestimate the energy of the flares unless we take into account the flux from the nearby star. Similarly, some flares may originate from the nearby star and be mistaken as flares from the target star. We determined the spectral type of these using the {\tt SIMBAD} database and seven of them are classified as M dwarfs.  Therefore, we cannot exclude that a small fraction of flares from these stars could originate from the nearby star. We come back to this issue in \S 7 where we investigate the rotational phase distribution of the flares.

Additionally, it is important to mention the differing band-passes between Kepler and TESS. The TESS band-pass spans from 600 -- 1000~nm and is centred on 786.5~nm \citep{Ricker2015}, whereas the Kepler band-pass spans 400 -- 900nm. This redder band-pass was chosen specifically to observe a larger number of M dwarfs as planets are easier to detect around these smaller, cooler stars. This implies the TESS band-pass is more sensitive to redder wavelengths, therefore, TESS will not detect the less energetic events as flares from M dwarfs typically have their peak emission towards the blue.

\begin{figure*}
    \centering
    \includegraphics[width = 1.00\textwidth]{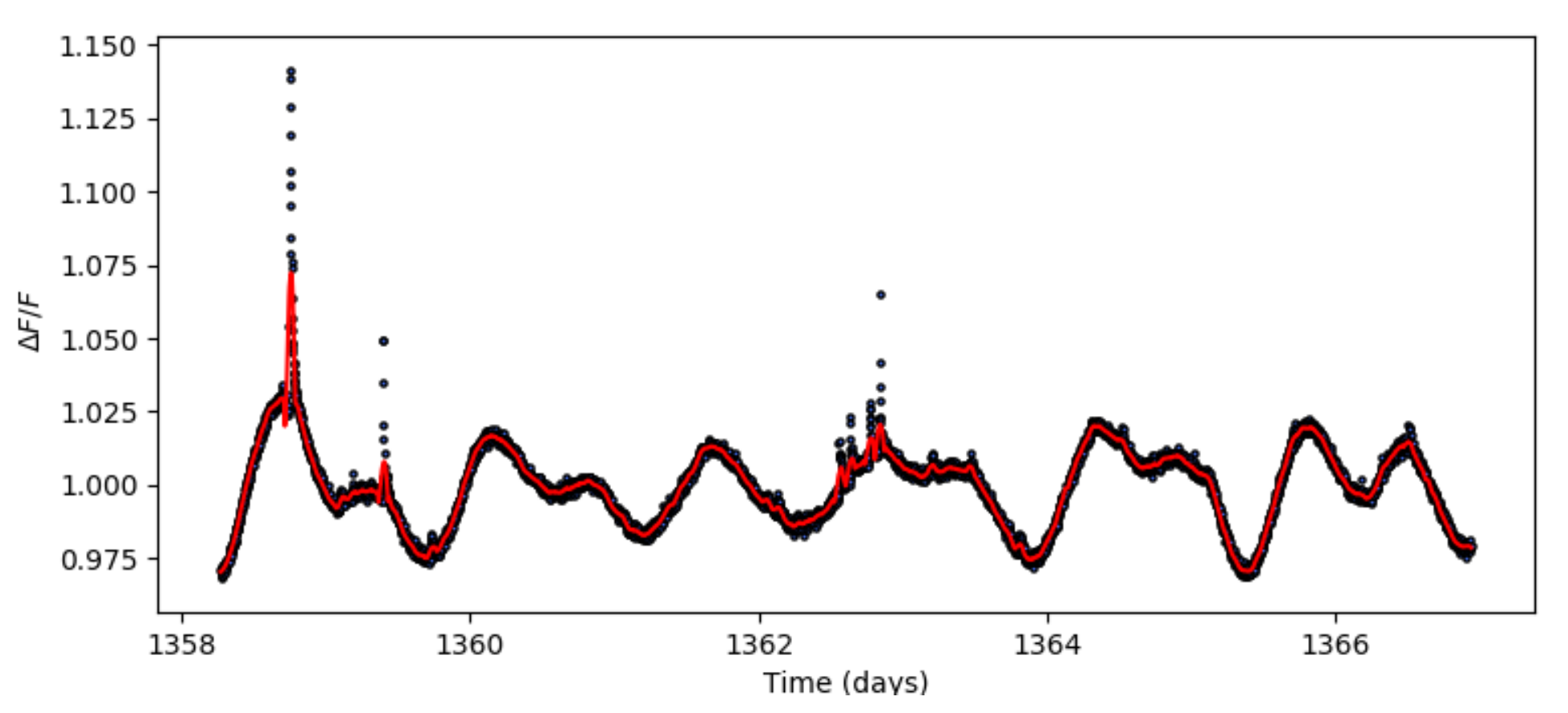}
    \caption{A section of the TESS lightcurve for 2MASS J0030-6236 (TIC 231914259) from Sector 1 which covers $\sim$ 9 days. This star has a spectral type, M2V and rotation period, $P_{rot}$, of  1.43 days. The black points represent the TESS data points which have a cadence of 2 mins and the red line is the Savitzky-Golay filtered, smoothed data and shows evidence of multiple spots and flares of varying magnitudes. It is important to note the rotational modulation within this particular star suggests the presence of two starspots. }
    \label{lightcurve_sec}
\end{figure*}

\section{Tess Data}

The first batch of TESS data was released in late 2018/early 2019 and includes observations from Sectors 1--3 made between July 25th and October 17th 2018.  The data release includes both Full Frame Images (FFIs) and Short Cadence (2 min) lightcurves. The 2 min lightcurves allow for the detection of short duration, low amplitude flares, and it is these lightcurves which form the basis of this work. We downloaded the calibrated lightcurve for each of our target stars from the MAST data archive\footnote{\url{https://archive.stsci.edu/tess/}}. We used the data values for {\tt PDCSAP\_FLUX}, which are the Simple Aperture Photometry values, {\tt SAP\_FLUX}, after the removal of systematic trends common to all stars in that Chip. Each photometric point is assigned a {\tt QUALITY} flag which indicates if the data may have been compromised to some degree by instrumental effects. We removed those points which did not have {\tt QUALITY=0} and normalised each lightcurve by dividing the flux of each point by the mean flux of the star. 
Each lightcurve was initially examined by eye to determine whether there was evidence for a rotational modulation and any flare-like events. Some sources which showed complex lightcurves (such as BY Dra variables) were not considered further. Additionally, we did not include the 18 stars which showed no modulation but did show flares in their TESS lightcurve as we could not determine the rotational phase of the flares, see Appendix \ref{no_mod_flares}. This leaves us with a final sample of 149 M dwarfs stars remaining in our sample for this study.

\section{Rotation Period}

Large variations in brightness can be observed in the lightcurves of M dwarfs which are widely explained by the presence of spots \citep{olah1997time}. These large, dominant starspot(s) come in and out of view as the star rotates producing quasi-sinusoidal changes in brightness, see Figure \ref{lightcurve_sec}. Observations with high enough cadence and length present one way of determining a stars rotation period. Thousands of low mass stars now have derived accurate rotation periods through Kepler and K2 observations  \citep[e.g.][]{mcquillan2014rotation}.

For each of the stars in our sample we determine the rotational period, $P_{rot}$, initially using the {\tt Period} \citep{dhillon2001period} software package which runs a Lomb-Scargle (LS) periodogram, followed by an iterative process involving the phase folding of sections from the start and end of the lightcurve. Phase zero, $\phi_{0}$, is also defined as the minimum of the flux of the rotational modulation which is initially determined by eye. The iterative process allows us to fine tune $P_{rot}$ and $\phi_{0}$ to fit the start and end of the data which ultimately derives the best fit to the data as a whole. We estimate the error on the period as a few percent which ensures the phase of all the flares are reliable. 

In Table \ref{stellar_properties} we show the stellar properties of our sample including $P_{rot}$ and $\phi_{0}$. The rotation periods of our sample range from 0.1 to 17.4 days. This includes 9 ultra-fast rotators with $P_{rot} < 0.3$ days and 53 with $P_{rot} < 1$ day. For stars with other stars spatially nearby, there is a small possibility in which the variability is due to the nearby star. 

\begin{figure}
\centering
\includegraphics[width = 0.47\textwidth]{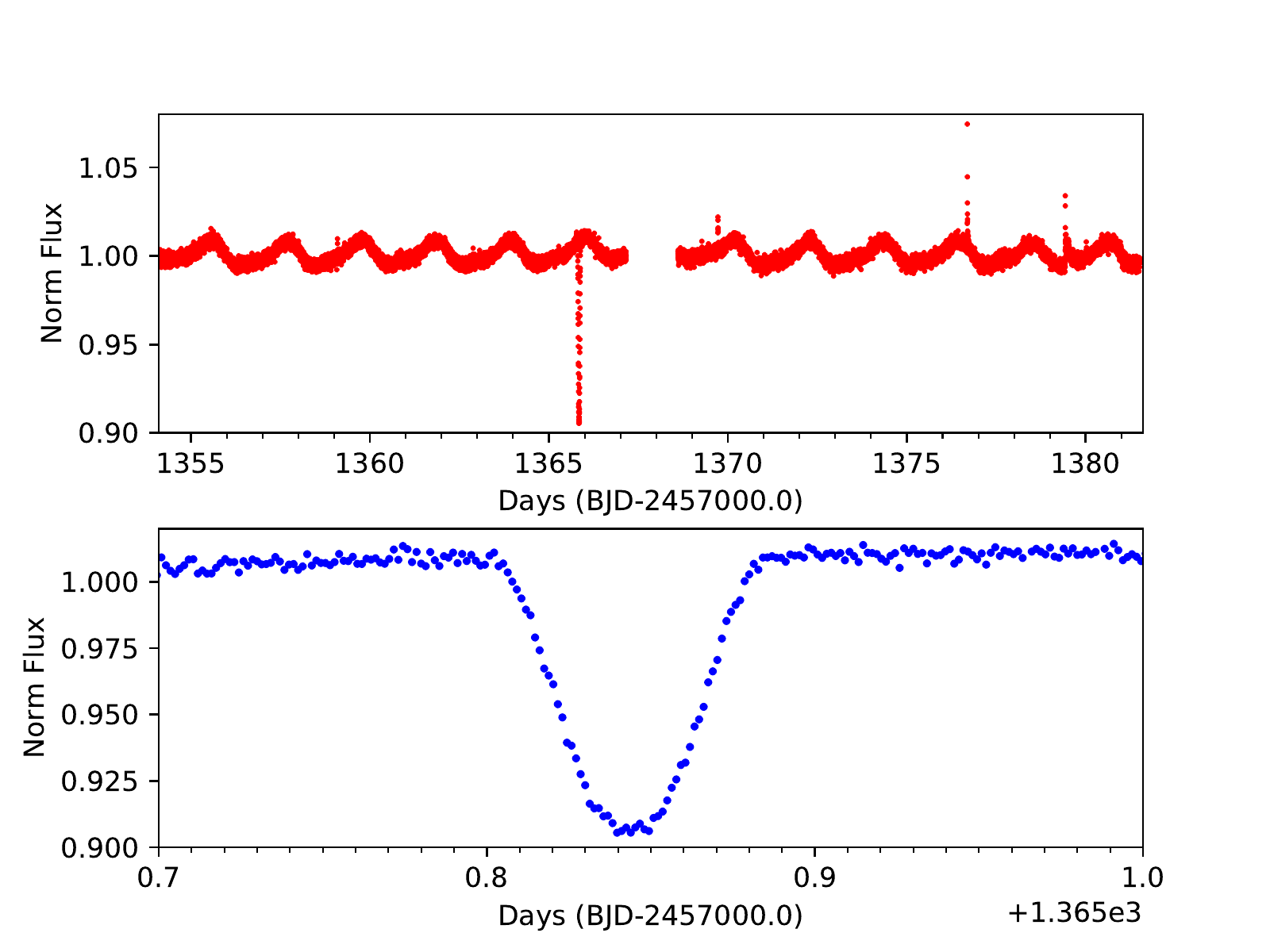}
\caption{The TESS short cadence data of G267-34 obtained in Sector 2 (top panel) where the flux has been normalised; the lightcurve zoomed in on the eclipse (lower panel). There is some evidence for asymmetry in the eclipse profile near mid-eclipse. In principal, observations like these can be used to map the distribution of the starspots \protect\cite[see][]{silva2003}. The lightcurve of this star (top panel) is an excellent example of stable rotational modulation as a result of one starspot present on the stellar disk.}
\label{eclipse}
\end{figure}

G 267-34 (M3V) shows a modulation on a period of $\sim2.1$ d and also a partial eclipse lasting $\sim$0.09 days with a depth of $\sim$0.1 (Figure \ref{eclipse}). Based on the known parameters of the star and the eclipse characteristics we estimate the second body is a late type M dwarf. Additionally, we only have three sources with rotation periods, $P_{rot} > 10$ days, in comparison to half of the sample from Paper I. This is entirely due to the observation length of TESS which is 27 days per Sector, making it difficult to observe slow rotating stars. 

\vspace{5mm}

\section{Stellar Flares}

\begin{figure}
    \centering
    \includegraphics[width = 0.47\textwidth]{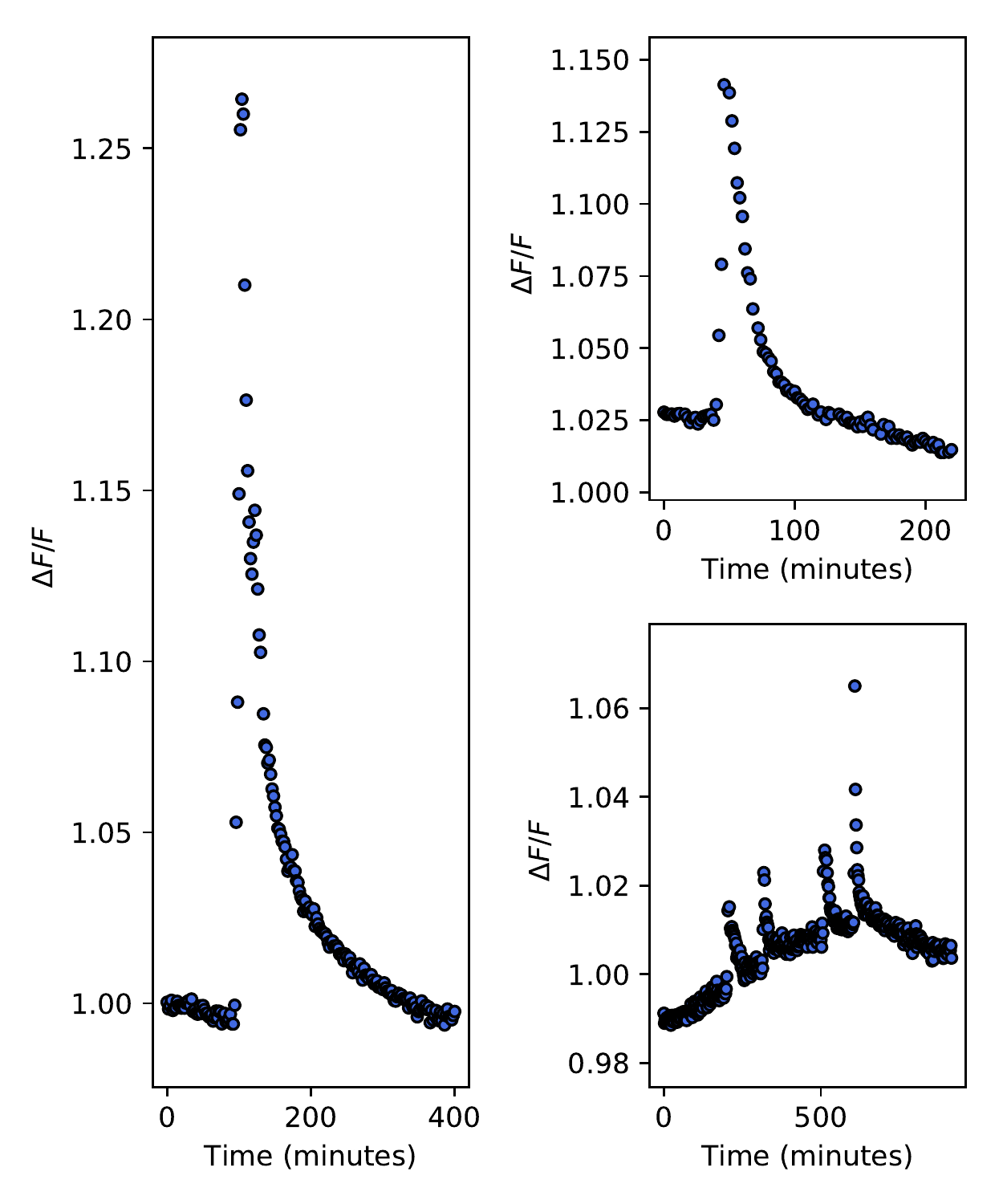}
    \caption{A selection of flares of varying magnitudes and duration from the M2.2 dwarf 2MASS J0030-6236 (TIC 231914259). This star was observed in Sectors 1 \& 2 for a total duration of $\sim$ 54 days, has a rotation period, $P_{rot}$, of 1.43 days and a total flare number of 58. The far left panel shows the largest flare from this star which a normalised flux peak of $\sim$ 1.27 and the remaining panels contain flares with lower energies.}
    \label{flare_image}
\end{figure}

For each star in our sample we want to identify the flares present in each lightcurve along with their energies in the TESS band-pass. Our approach to identifying flares is the same as Paper I. We use {\tt FBEYE} \citep{davenport2014kepler} which scans each lightcurve and flags up any point which is over a $2.5\sigma$ threshold, identifying potential flares which consist of 2 or more consecutive flagged points. Once complete, this produces a comprehensive list of flares per star along with properties for each flare such as start and stop time, flux peak and equivalent duration. A selection of flares of varying magnitudes can be seen in Figure \ref{flare_image} from the M2V dwarf 2MASS J0030-6236 (TIC 231914259) which has a rotation period, $P_{rot}$ = 1.43 days. Overall, we catalogue 1765 flares from the 149 flaring dwarf stars with a range of magnitudes and durations (see Table \ref{stellar_properties}). For stars which showed no modulation we catalogue 69 flares and show them in Appendix \ref{no_mod_flares}.

\begin{figure}
    \centering
    \includegraphics[width = 0.47\textwidth]{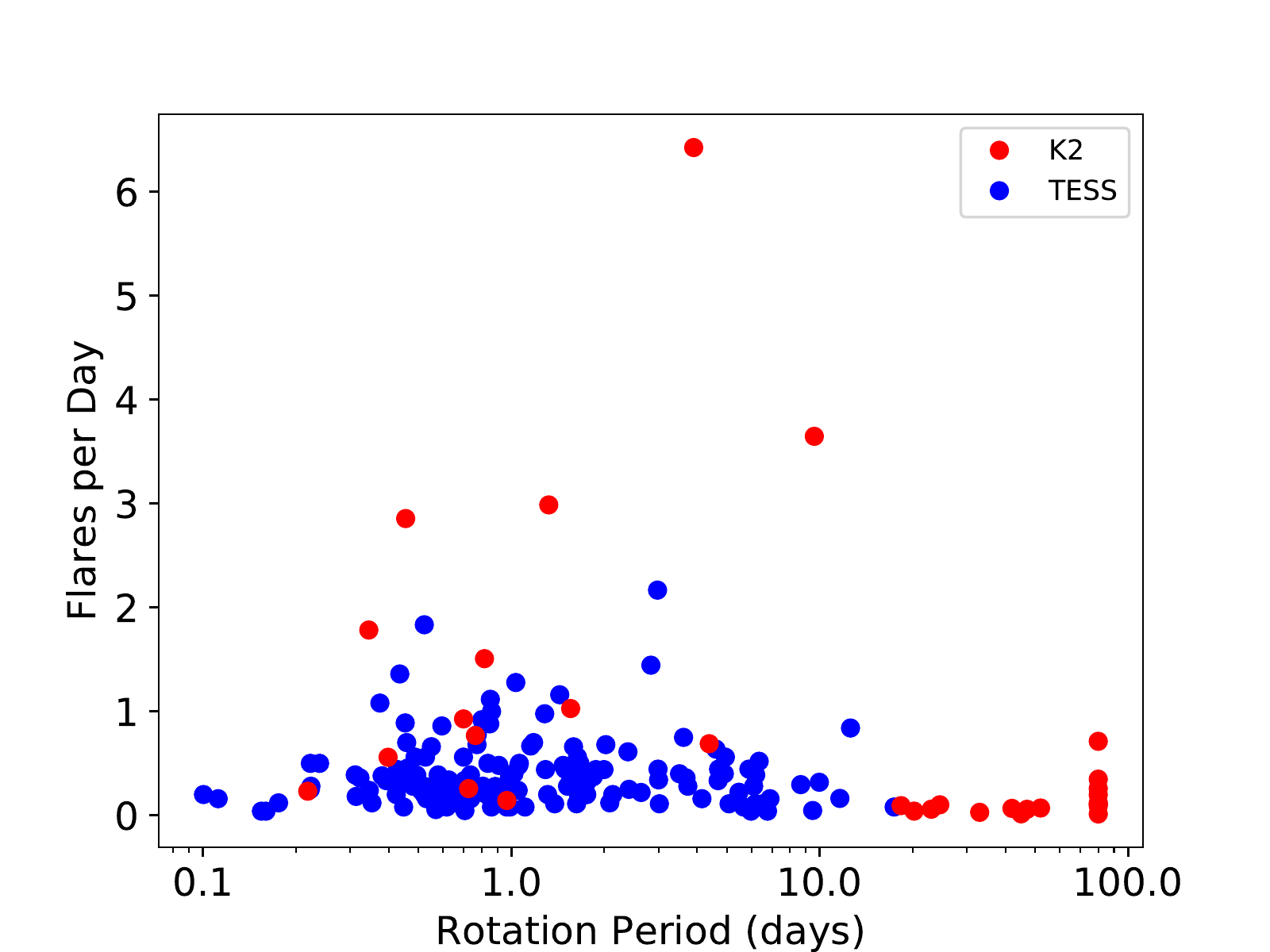}
    \caption{Here we show the normalised number of flares per day for each star as a function of rotation period. The red represent the stars from Paper 1 using K2 short cadence data, and the blue the 149 targets from this study using TESS 2-min cadence data.} 
    \label{rot_vs_flareno}
\end{figure}

Figure \ref{rot_vs_flareno} shows the normalised flare number per day for each star in our sample according to the stellar rotational period. We also overplot the M dwarfs from Paper I  which were observed using short cadence K2 data. As we found in Paper I, we see a drop in flare number for stars with $P_{rot} > 10$ days, however, our sample is limited in this respect due to the observation length of TESS being 27 days for each Sector. On average the flare rate of the TESS stars is lower than that of the K2 stars which we attribute to its lower sensitivity. 

The energies of the flares are determined as the equivalent duration of each flare multiplied by the quiescent luminosity of the star. Amongst our sample of 149 M dwarfs, a wide range in flaring energies are observed. The lowest energy flare is $\sim 6\times 10^{29}$~erg and is observed in the M3 star PMJ 01538-149 (TIC 92993104) from Sector 3 with $P_{rot} = 3$ days. Similarly, the highest energy flare is $\sim 2\times 10^{35}$~erg from an M3 star GSC 08494-00369 (TIC 201897406) observed in Sector 2 with $P_{rot} = 3.5$ days. All of the properties including the luminosities of the stars and range of flare energies can be found in Table \ref{stellar_properties}. We should caution that for stars with spatially nearby stars these luminosities are underestimated by up to a factor of two, see the online table for a list of those objects. 

\begin{table*}
\caption{The stellar properties of a select few stars in our survey detailing the number of flares, rotation periods, quiescent luminosity, energy range and duration range of the flares. The apparent magnitude in the TESS band-pass, $T_{mag}$, is taken from the TESS Input Catalog (TIC) along with the TIC ID \citep{stassun2018tess}. The distances are derived from the Gaia Data Release 2 parallaxes \citep{gai16, gaia18} and the spectral types are obtained from the {\tt SIMBAD} catalogue. \newline
{\it This table is available in its entirety in a machine-readable form in the online journal. A portion is shown here for guidance regarding its form and content.}}

   \begin{center}
   \label{stellar_properties}
\resizebox{\textwidth}{!}{
	\begin{tabular}{lccccccccccc}
    \hline 
	Name                  & TIC ID     &  Sector     & No. of   & SpT    & $P_{rot}$ & $T_{mag}$  &  Parallax             &  Distance             &  $log(L_{star})$  & $log(E_{flare})$   &  Duration        \\
	                      &            &             & Flares   &        & days      &            &  mas                  &  pc                   &  erg/s            & erg             &  minutes         \\
	\hline
	UCAC4 110-129613      & 229807000  &  1          & 27       &  2.5   & 0.3745    & 10.737     & $21.716 \pm 0.0235$   & $46.0481 \pm 0.0498$  &  31.79            & 31.59 -- 34.31  & 8.00 -- 90.00    \\
    WOHS 209              & 179038379  &  1          & 5        &  0.0   & 11.6169   & 12.621     & $16.130 \pm 0.0434$   & $61.9959 \pm 0.1668$  &  31.28            & 32.44 -- 33.91  & 16.00 -- 134.00  \\
    UCAC3 53-724          & 425937691  &  1\&2       & 10       &  5.5   & 0.1003    & 13.178     & $22.794 \pm 0.1100$   & $43.8708 \pm 0.2117$  &  30.80            & 32.15 -- 33.54  & 10.00 -- 60.00   \\
    CD-561032B            & 220433364  &  2          & 37       &  4.0   & 0.8543    & 9.380      & $90.165 \pm 0.0379$   & $11.0907 \pm 0.0047$  &  31.10            & 31.58 -- 33.45  & 8.00 -- 72.00    \\
    WISE J0305-3725       & 165124012  &  3          & 8        &  1.9   & 2.9880    & 10.480     & $5.465 \pm 1.508$     & $182.952 \pm 50.498$  &  32.90            & 33.49 -- 34.68  & 12.00 -- 78.00   \\
    CD-52381B             & 229151691  &  3          & 7        &  9.5   & 2.4219    & 9.890      & $11.587 \pm 0.040$    & $86.2977 \pm 0.2994$  &  32.66            & 33.61 -- 35.29  & 26.00 -- 309.99  \\
    \hline
    \end{tabular}}
    \end{center}
\end{table*}

\section{Rotational Phase}

For all stars in our sample, they display a clear rotational modulation which we attribute as being due to the presence of a large dominant star spot rotating into and out of view. From studies of solar flares we know numerous and more energetic flares of energies ($\geq 10^{32}$~erg) occur in active regions which possess complex sunspot configurations \citep{zirin1982delta, mcintosh1990classification}. We would therefore expect to observe the same behaviour from M dwarfs which host a large, dominant spot as part of a complex active region. In Paper I we investigated the preference for rotational phase for a small group of M dwarfs observed in short cadence by K2 and found no correlation between flare number and rotational phase, despite the clear presence of large dominant starspots. For this sample of M dwarfs observed using TESS we used the same analysis as Paper 1 to determine if any of the flares show a preference for certain rotational phases which coincide with the dominant starspot.   

We use the $\chi_{\nu}^2$ statistic as a means of assessing the rotational phase distribution of the flares. In order to do this we split our sample depending on the number of flares present in the lightcurves of each star using the overall mean number of flares as an indicator. Any star which possessed >~12 flares in its TESS lightcurve was considered to be an active M dwarf and remaining sources are grouped together for the rotational phase analysis of the flares. We also look at the flares from all 149 low mass dwarf stars as a whole and present and discuss these results individually.

\subsection{Individual Cases}

\begin{figure}
    \centering
    \includegraphics[width = 0.47\textwidth]{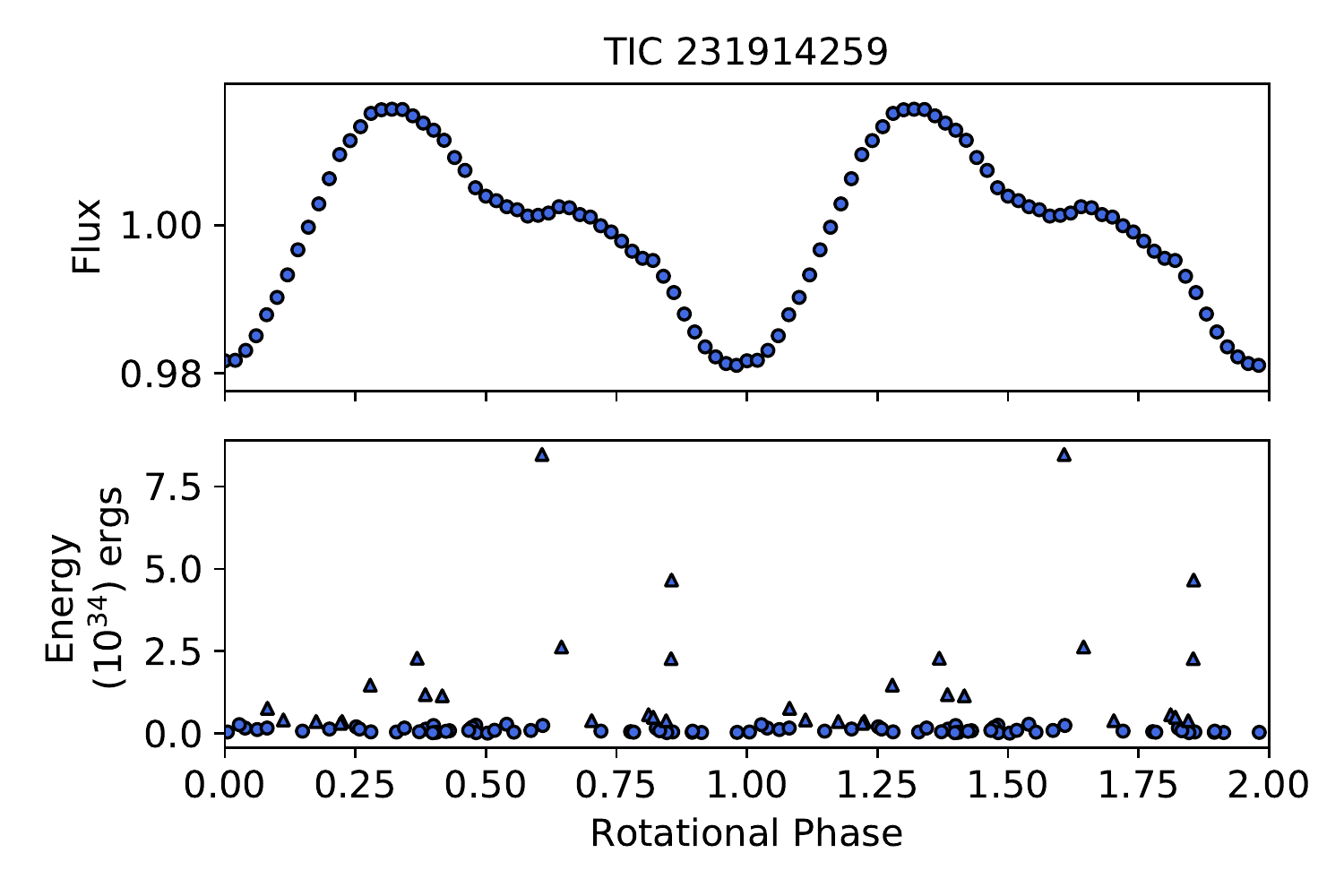}
    \caption{The rotational phase distribution for 2MASS J0030-6236 (TIC 231914259) observed in Sectors 1 \& 2 (where we repeat the rotational phase coverage $\phi=0.0 -2.0$). The upper panel shows the phase folded, binned lightcurve where phase zero is defined as flux minimum and $P_{rot}$ = 1.43 days. The lower panel shows the  phase distribution of the flares as a function of energy where triangle symbols represent flares of energies > $3.16\times10^{33}$erg and circles < $3.16\times10^{33}$erg.}
    \label{phase_energy}
\end{figure}

From our sample, 45 stars show 13 or more flares in their lightcurves so, we can test the phase distribution of these flares individually. Out of this 45, ten have a star which was up to 1.5 mag fainter and within 42$^{''}$ of the target. For each of these stars we phase fold and bin their lightcurves using rotation periods and phase zeros obtained previously. The flares are then split into low and high energy, where the cut-off was determined from a histogram distribution of all flares which levelled off at $10^{33.5}$~erg. This information can be displayed in plots similar to Figure \ref{phase_energy} which show the rotational phase distribution of the flares as a function of energy. In this particular example, 2MASS J0030-6236 (TIC 231914259) shows a total of 58 flares, 17 of which are considered to be high energy, which are spread across all rotational phases. 

A simple $\chi_{\nu}^2$ test is used to assess the randomness within the rotational phase distribution of the flares. The rotational phase, $\phi$, is split into 10 bins (degrees of freedom is 9) of 0.1 between $\phi = 0.0 - 1.0$ and $\chi_{\nu}^2$ is calculated for high energy, low energy and all flares. As an example, 2MASS J0030-6236 (Figure \ref{phase_energy}) has $\chi_{\nu}^2$ of 1.31, 1.32 and 1.48 for high, low and all flares: none are significant. For full details of this method please refer to Paper I. Overall, out of the 45 M dwarfs none show flares which have a preference for rotational phase indicating the flares are not associated with a large, dominant starspot. 

\subsection{The Remaining Sources}

\begin{figure}
    \centering
    \includegraphics[width = 0.47\textwidth] {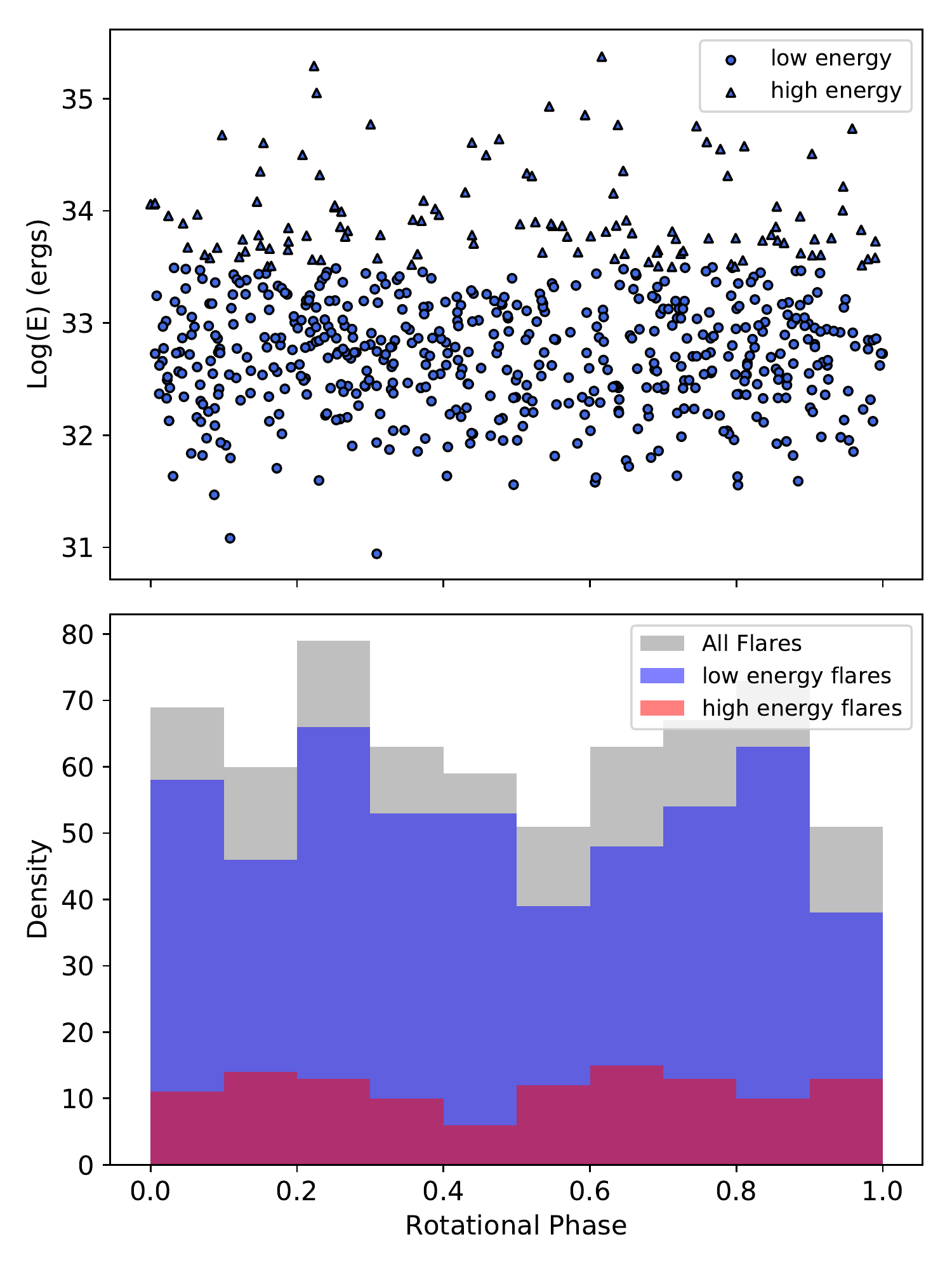}
    \caption{The rotational phase distribution for all stars which show $<$ 12 flares in their TESS lightcurves, where $\phi$ is defined as flux minimum 0.0 which represents rotational minimum. The upper panel shows the rotational phase distribution as a function of energy where triangle symbols represent flares of energies >~$3.16\times10^{33}\;\!$erg and circles <~$3.16\times10^{33}\;\!$ergs. The lower panel shows the histogram of the rotational phase distribution.}
    \label{remaining_all}
\end{figure}

For the 104 remaining M dwarfs in our sample which show $\leq$~12 or less flares in their lightcurves, we group them together to asses the rotational phase distribution of the flares as a whole. As we define phase zero, $\phi_{0}$, for each star to be the minimum of the rotational modulation, making a comparison of the flare phase distribution for varying stars possible. Figure \ref{remaining_all} shows the flare rotational phase distribution as a function of energy for all flares from the remaining 104 stars, along with the histogram of the distribution where the rotational phase has been split into 10 bins of 0.1 between $\phi = 0.0 - 1.0$. Flares were also split into low and high energy using the cut off mentioned previously of $10^{33.5}$ erg. We applied the $\chi_{\nu}^2$ test to this sample of flares with values of 1.25, 0.57 and 0.57 for all, high and low flares respectively. Again, these $\chi_{\nu}^2$ values for high and low energy flares indicate there is no correlation between rotational phase and flare number amongst this group of flares from 104 stars.

This was repeated splitting the stars into spectral type categories of < M4 and > M4 with no preference for flare rotational phase found in either of these groups. Similarly, the stars were split by rotational period, for example 0~-~0.5 days, 0.5~-~1 days and so on, with no evidence of any correlation being found using the $\chi_{\nu}^2$ test. In addition, we also removed all targets with nearby stars, according to Gaia, finding values of the $\chi_{\nu}^2$ test to be 1.13, 0.945 and 0.945 for all, high and low flares respectively: our conclusions therefore are not affected.

To summarise, no evidence was found of any correlation between flare number and rotational phase in the grouped remaining 104 M dwarfs which show $\leq 12$ flares in their lightcurves. When splitting this group up by spectral type or rotational period again no correlation was found and the $\chi_{\nu}^2$ statistical test indicated the flares are randomly distributed.

\subsection{The Sample as a Whole}

\begin{figure}
    \centering
    \includegraphics[width = 0.47\textwidth]{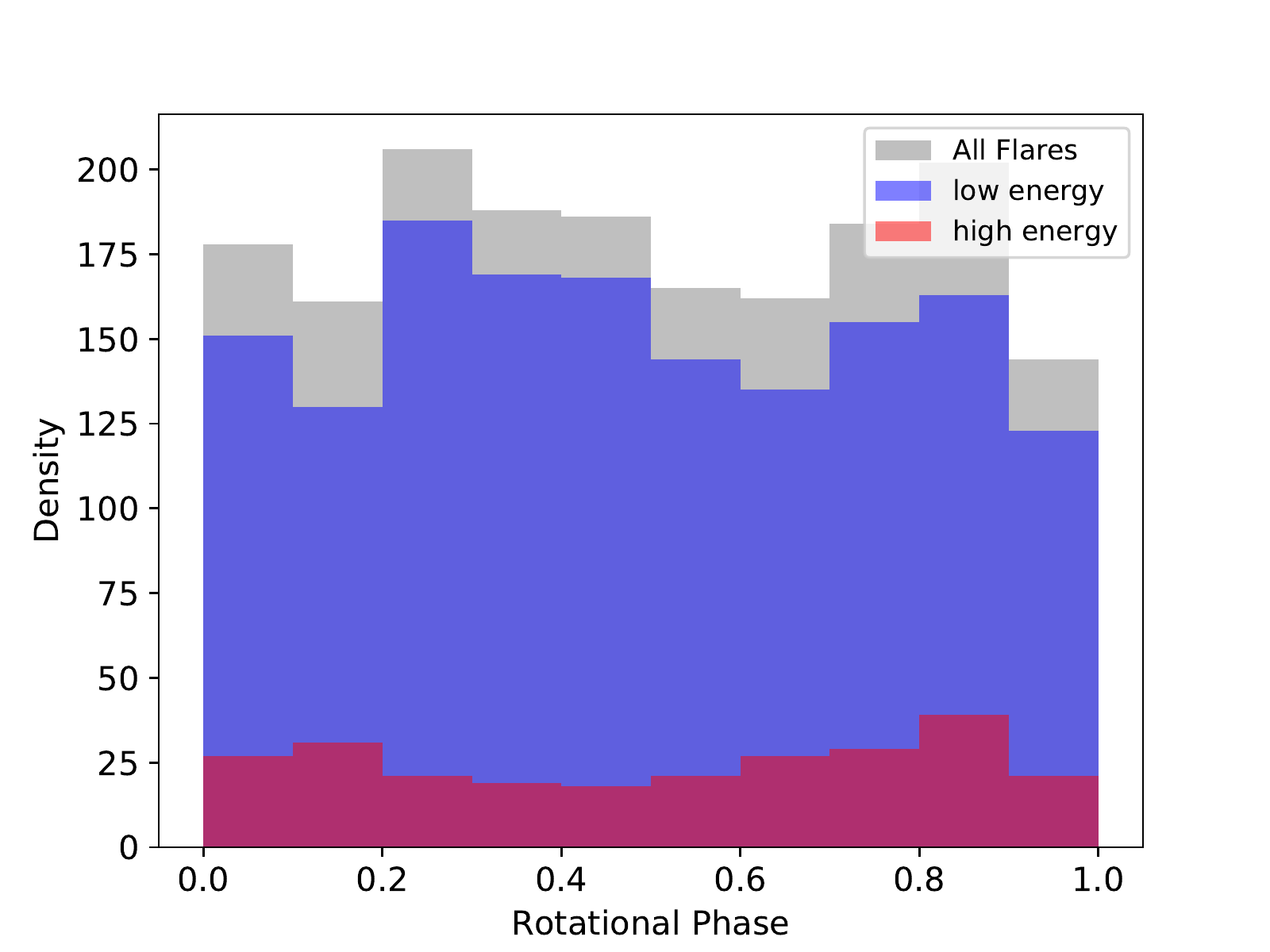}
    \caption{The rotational phase distribution for all 1776 flares from the sample of 149 stars. We show the histogram of this distribution using bins of $\phi = 0.1$ where there is no evidence of any correlation between flare number and rotational phase in high, low or all flares. }
    \label{all_phase}
\end{figure}

We also look at the rotational phase distribution of all 1765 flares from the sample of low mass stars to check for any correlations with flare number. As $\phi_{0}$ is defined at flux minimum we are able to make this comparison possible. We do not find any correlation between flare number and rotational phase and the $\chi_{\nu}^2$ test indicates the flares are randomly distributed. Figure \ref{all_phase} shows the histogram distributions of the rotational phase for all flares which displays a uniform spread amongst the phase bins of $\phi = 0.1$. This was repeated for phase bins of $\phi = 0.2$ and $\phi = 0.01$ and again no correlation was found. In addition to the $\chi_{\nu}^2$ test we also looked at testing the distribution of flares using the Kolmogorov-Smirnov (KS) and  Shapiro-Wilk (SW) tests. The results were, KS  = 0.5 with p-value = 0 and SW = 0.55 and p-value = $7.3\times 10^{-23}$. These results suggests the data conforms to a normal distribution and therefore, according to the Central Limit Theorem is a random distribution. This analysis provides consistent results to the $\chi_{\nu}^2$ test even with the removal of the bin dependence. 

This analysis is similar to that of \cite{roettenbacher2018connection} where they look at the number of flares occurring in bins of phase $\phi = 0.01$, for all stars in their sample of 119. They do find a correlation between rotational phases which represent a visible starspot and flare number. This is observed as a peak in the histogram distribution for flares of flux increases between 1\% and 5\% only and not in higher energy flares of flux increases $>\;\!$5\%. However, their sample differs greatly from ours as it consists of main sequence stars from late-F to mid-M observed in long cadence (30 min) by Kepler. 

Similarly, in \cite{roettenbacher2018connection} they only select stars for their sample which exhibit one spot structure at a time in their lightcurve. From our sample of 149 M dwarfs which show rotational modulation in their lightcurves, 77\% show a clear sinusoidal pattern indicative of the presence of one large starspot, where the remaining 23\% have lightcurves which would have arisen from multiple spots. To address this, we removed all stars which showed any evidence for multiple spots within its lightcurve: our conclusions do not change.

\section{Ultra-Fast Rotators}

Rapidly rotating low mass stars are expected to produce increased levels of activity which is strongly related to their dynamo mechanism \citep{hartmann1987rotation, maggio1987einstein}. Activity is saturated in rapid rotators with a decline in activity as a function of rotation \citep{kiraga2007age,yang2017flaring}. 

\begin{figure}
    \centering
    \includegraphics[width = 0.47\textwidth]{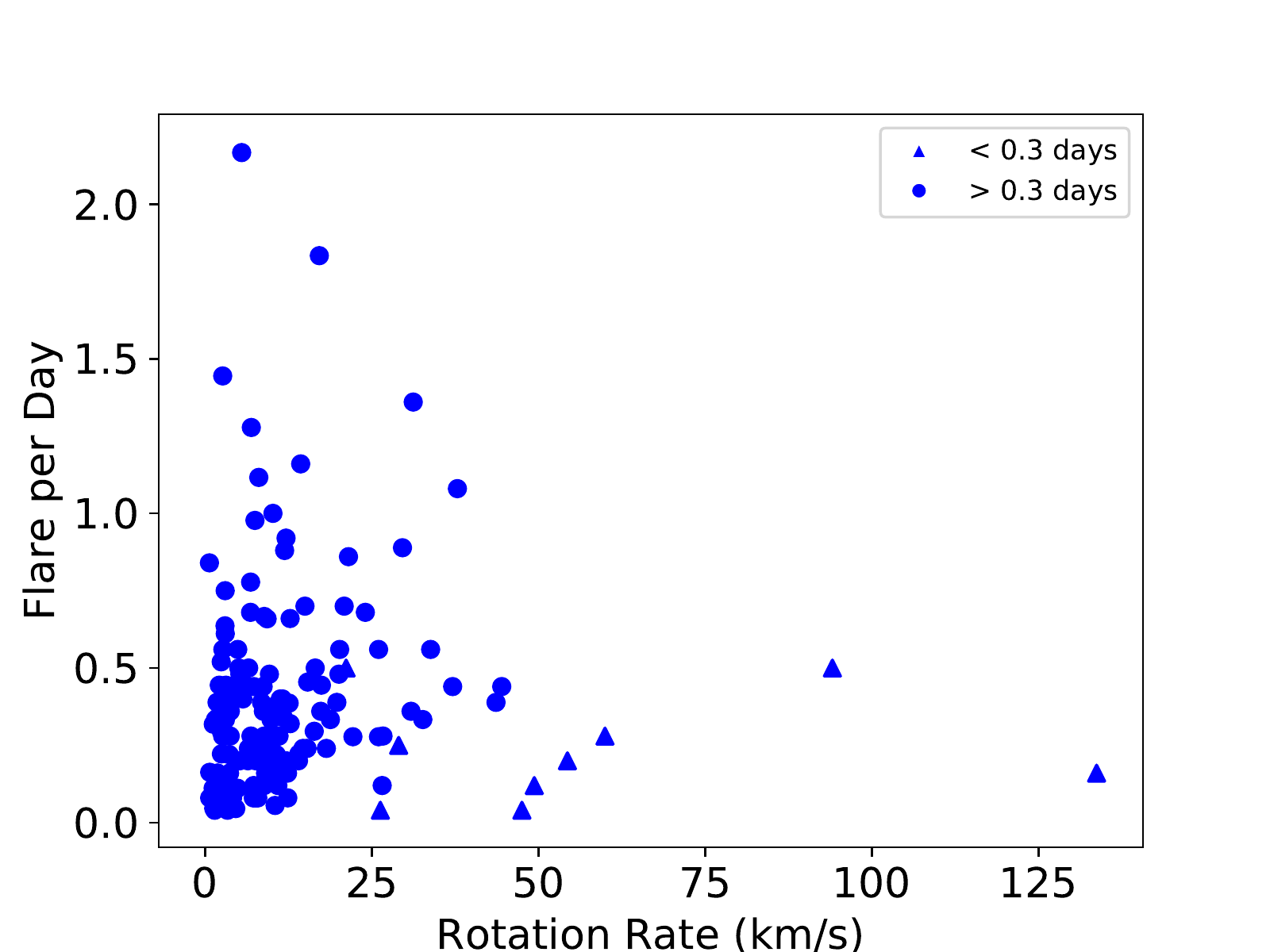}
    \caption{The normalised flares per day of each star in the TESS sample as a function of rotation rate.}
    \label{rotation_rate}
\end{figure}

From Figure \ref{rot_vs_flareno}, we notice a small group of nine stars which have rotation periods, $P_{rot} < 0.3$ days and spectral types in the range from M1 to M6 which produce a very low number of flares. To investigate this further we determined the rotational velocity, $\Omega$, for all stars in our sample as $\Omega = 2\pi R/P_{rot}$ which has units of $km/s$. $R$ is the radius of the star derived from the Stefan-Boltzmann Law using the temperature derived from Gaia DR2 and $P_{rot}$ is the rotation period. Although there is some uncertainty in determining $\Omega$, Figure \ref{rotation_rate} shows that the stars with $P_{rot} < 0.3$ days also have high rotational velocities. Secondly, we are surprised to find that the flare rate decreases with increasing rotational velocity. We would expect the stars with higher rotational velocity to show greater flaring activity. At this point we are unable to explain this finding which will therefore require further investigation.

We look to determining stellar ages for the small number of ultra-fast rotators as another indicator to explain the lack of flaring activity. To do this we use Gyrochronology \citep{barnes2007ages} which uses a relationship between the age, colour and rotation period of main sequence stars. For the purposes of this work we use {\tt stardate} \citep{2015isochrones, 2019stardate} a python package which combines isochrone fitting with gyrochronology. For the nine ultra-fast rotators we find they have ages estimated to within the range of 3~Myr to 2~Gyr. If we take these ages at face value it raises serious questions for how such a rapid rotator could be as old as 2 Gyr, and conversely how such a young fast rotator can show only few flares. However, although the uncertainties in determining the ages of solar-type stars is reasonably well understood, the spread in period against age for low mass stars is much higher. With this caveat in mind, we conclude that age may not be the primary cause for the lack of flaring activity in these stars. Rather, this suggests it may be related to their magnetic field configuration. 

\cite{kochukhov2017global} recently investigated the global and small-scale magnetic field configuration of the nearby M dwarf binary GJ65 AB. Despite nearly identical masses and rotation rates ($P_{rot} = 0.2432$ days for GJ65 A and $P_{rot} = 0.2269$ days for GJ65 B  \citep{barnes2017surprisingly}), the secondary exhibits an axisymmetric, dipolar-like global field with an average strength of 1.3 kG while the primary has a much weaker, more complex, and non-axisymmetric 0.3 kG field. Furthermore, GJ 65 B flares more frequently than GJ 65 A and is also an order of magnitude brighter in its steady radio emission \citep{audard2003separating}. Despite their rotation rate both of these stars possess dramatically different magnetic field configurations along with varying degrees of magnetic activity. This suggests the magnetic field configuration of the stars plays an important role in their magnetic activity, more so than their rotation period or age. However, it is likely that all three factors are connected. 
 
 An alternative explanation is that flares from these objects emit mostly in the blue. We have made a preliminary check for a few known ultra-fast rotators (KIC 6752578, KIC 6791060 and KIC 9825598) from literature observed by Kepler and find they too show a low number of flares. However, these sources have only been observed in long cadence (30 min), compared to TESS SC at 2-min, which could be the reason for not observing the short duration flares if they are present. Additional work is needed via checking the flare activity on the same object as observed by Kepler and TESS. Therefore, this must wait until observations begin in the northern hemisphere with TESS where it will overlap with the Kepler field. 

\section{Discussion}

We have analysed the magnetic activity of a sample of 149 low mass stars observed in 2-min cadence by TESS. We have successfully derived rotation periods for 90 percent of stars in our sample as a result of rotational modulation present in their lightcurves. In addition, we have catalogued the flare characteristics (energy, duration, rotational phase, etc.) of each star and use this information to compute a statistical analysis on the flaring activity. We find no evidence of a correlation between rotational phase and flare number for any star in our sample or all 149 stars as a whole. Even when we remove targets with nearby stars there is still no trend present with regards to rotational phase and flare number. Furthermore, when we restrict our sample to only those which show evidence of one starspot we still find no evidence for a correlation between rotational phase and flare number. This result is consistent with our findings in Paper I and with a larger sample size this solidifies our initial finding. In Paper I we proposed three scenarios to explain the lack of correlation between phase and flare number including binarity, presence of exo-planets and polar spots. Here we will discuss these three scenarios in greater detail while also introducing a number of others which have come to light since Paper I. 

Firstly, we discuss the possibility of star-star and star-planet interactions being a cause of the constant flaring activity at all rotational phases observed in our samples of M dwarfs from both K2 and TESS. In \cite{fischer2019time} they look to identify signatures of star-planet interaction (SPI) in the K2 lightcurve of the TRAPPIST-1 system. They discuss four mechanisms which cause temporal variability of SPI  lightcurves, two by orbital positions of planets and two due to the stellar magnetic field. Overall, their results hint at a possibility of a quasi-periodic occurrence of flares with the orbiting planet TRAPPIST-1c. However, this result is inconclusive due to various factors but is a promising potential find for further studies. Similarly, \cite{route2019rise} use multi-wavelength observations to study the SPI in the HD~189733 system. The star in this system is a very active BY~Dra type variable with a Jupiter mass planet orbiting at a distance of 0.031 AU. Through physical and statistical analysis, \cite{route2019rise} conclude there is no existence of SPI within this system and stellar activity on HD 189733A is not correlated at certain orbital phases.

In earlier work, \cite{van1998prominence} and \cite{byrne1998evidence} showed that for a star with radius $\sim$0.3R$_\odot$, a field strength of $\sim$2.5kG (typical values for an M dwarf), the maximum amount of stored energy is $\sim10^{37}(l/R_\odot)$ erg where $l$ is the length of a filament and $R_\odot$ is the solar radius. Thus for a filament whose length is $\sim$30\% of the star's radius this equates to $\sim10^{36}$ erg. This is more than sufficient to explain the large flare energies for flares on M dwarfs. Regarding the possibility of a filament located between the star and a nearby planet this allows a factor of $(1.6a/R_*)^2$, where $a$ is the binary separation and $R_*$ is the star's radius. However, despite these studies SPI is still a relatively new area of research and is difficult to observe. Hopefully further observations with missions like TESS will provide the observations needed to prove SPI.  

As mentioned previously, \cite{roettenbacher2018connection} conducted a similar study to investigate the connection between starspots and flares in main-sequence stars. They use a sample of late-F to mid-M stars observed using long cadence (30 min) lightcurves over 4 years. From their sample of 119 stars, 2447 flares were detected with only lower energy flares occurring predominantly with the large starspot. They propose this could be a result of more energetic flares being observed on disk and close to/over the limb, whereas less energetic flares would not be strong enough to be seen over the limb. In the present data we do not see any difference in the rotational phase versus either the large or small flares. Furthermore, \cite{mariska1999hard} reported that in most instances, occulted solar limb flares were indistinguishable from non-occulted limb flares, although the hard X-ray spectra averaged over the entire event had a softer spectral index in the occulted limb flares thus indicating an occultation of the hard X-rays. \cite{kuhar2015correlation} shows a good correlation between hard X-ray fluxes and the excess white light flux. It is likely the white-light emission from occulted flares may not be observable, hence we may rule out this as an explanation for the lack of rotational modulation in the flare occurrence.  

An additional scenario includes the possibility of multiple spot locations across the disk of the star. Fitting a lightcurve with a one or even two spot model does not produce the sinusoidal pattern observed in many low mass stars from Kepler and TESS \citep{eaton1996random}. If the rotation modulation was a result of one dominant, large starspot, we would almost always observe flat-top lightcurves and this is not the case. Therefore, the sinusoidal pattern we do observe is not produced by a circular large starspot but in fact multiple active regions which hosts more/larger spots across the disk. One active region will possess either a larger spot or greater spot coverage and would be responsible for the peak and trough present in the sinusoidal lightcurve. In theory, this active region should produce higher energy flares, see \cite{mcintosh1990classification}. However, we do not observe any correlations between low or high energy flares and the phases corresponding to the minimum of rotational modulation. This suggests there are other magnetic features, such as polar spots, and/or magneto-kinetic/hydrodynamical processes at play.

Finally, there is the potential for the presence of polar spots on these low mass stars \citep{strassmeier1996observational}. Unlike the Sun, where polar spots are not present as a result of its dynamo mechanism, polar spots can be present in these low mass stars and depending on the viewing geometry are not reflected in the lightcurve. In Paper I, we discussed the formation of polar spots in more detail concluding they have the potential to play an important role in flare generation. It is possible they could be interacting with multiple spot groups across the disk to produce constant flaring activity at a range of energies and at all rotational phases. Similarly \cite{roettenbacher2018connection} also suggest the presence of polar spots as an explanation for the spread in stellar flares within their sample.

We have further discussed the three scenarios proposed in Paper I to explain the lack of a correlation between flare number and rotational phase. With our extended sample of 149 low mass stars observed in 2-min cadence with TESS, we solidify our initial finding and propose two new scenarios to explain its cause. However, it is likely the results we obtain stem from a combination of the five scenarios, in particular multiple spot locations and polar spots will have a big role to play. As a possible test of the polar spots theory, Zeeman-Doppler imaging with simultaneous flare monitoring of a flare star would be required. Such a dataset would enable a comparison of the spot location via Doppler imaging and the occurrence time of the flare. Flare monitoring could be done with e.g. TESS or Evryscope \citep{howard2019evryflare} which is an excellent facility for detecting the most energetic flares.

\section{Conclusions}

To summarise, we conduct a statistical analysis of stellar flares from a sample of 149 low mass stars observed in 2-min cadence by TESS. In particular, we focus on investigating the correlation between the rotational phase and number of flares and ultimately find no evidence of any such correlation. This is unexpected as you would expect a correlation in flare number to coincide with the minimum of rotational modulation when spot coverage is at its maximum. We therefore outline explanations for our finding including star-planet interactions, polar spots and multiple spot locations. 

At the time of writing, TESS is halfway through its initial 2-year primary mission and will soon be covering the northern ecliptic. Amongst these observations the Kepler field will be observed again providing a wealth of extended data. As TESS continues to observe the sky it will produce vast quantities of 2-min cadence data on thousands of low mass stars, ideal for the continuation of this study. 

We also touch upon a group of rapidly rotating stars within our TESS sample which have $P_{rot} < 0.3$ days but very little flaring activity. We speculate the reasoning behind this and conclude it is most likely to be a result of the magnetic field configurations of the star. In order to investigate this further we would look to obtaining spectropolarimetry observations to derive more information on the magnetic properties of the stars. 

\section*{Acknowledgements}

This paper includes data collected by the TESS mission. Funding for the TESS mission is provided by the NASA Explorer Program.
Armagh Observatory and Planetarium is core funded by the Northern Ireland Government through the Dept. for Communities. LD acknowledges funding from an STFC studentship. 
This work presents results from the European Space Agency (ESA) space mission {\sl Gaia}. {\sl Gaia} data is being processed by the {\sl Gaia} Data Processing and Analysis Consortium (DPAC). Funding for the DPAC is provided by national institutions, in particular the institutions participating in the {\sl Gaia} MultiLateral Agreement (MLA). The Gaia mission website is \url{https://www.cosmos.esa.int/gaia}. The Gaia archive website is \url{https://archives.esac.esa.int/gaia}. We thank the anonymous referee for their helpful report. 

The national facility capability for SkyMapper has been funded through ARC LIEF grant LE130100104 from the Australian Research Council, awarded to the University of Sydney, the Australian National University, Swinburne University of Technology, the University of Queensland, the University of Western Australia, the University of Melbourne, Curtin University of Technology, Monash University and the Australian Astronomical Observatory. SkyMapper is owned and operated by The Australian National University's Research School of Astronomy and Astrophysics. The survey data were processed and provided by the SkyMapper Team at Australian National University. The SkyMapper node of the All-Sky Virtual Observatory (ASVO) is hosted at the National Computational Infrastructure (NCI). Development and support the SkyMapper node of the ASVO has been funded in part by Astronomy Australia Limited (AAL) and the Australian Government through the Commonwealth's Education Investment Fund (EIF) and National Collaborative Research Infrastructure Strategy (NCRIS), particularly the National eResearch Collaboration Tools and Resources (NeCTAR) and the Australian National Data Service Projects (ANDS).

\bibliographystyle{mnras}
\bibliography{TESS_paper.bib} 

\begin{thebibliography}{}
\makeatletter
\relax
\def\mn@urlcharsother{\let\do\@makeother \do\$\do\&\do\#\do\^\do\_\do\%\do\~}
\def\mn@doi{\begingroup\mn@urlcharsother \@ifnextchar [ {\mn@doi@}
  {\mn@doi@[]}}
\def\mn@doi@[#1]#2{\def\@tempa{#1}\ifx\@tempa\@empty \href
  {http://dx.doi.org/#2} {doi:#2}\else \href {http://dx.doi.org/#2} {#1}\fi
  \endgroup}
\def\mn@eprint#1#2{\mn@eprint@#1:#2::\@nil}
\def\mn@eprint@arXiv#1{\href {http://arxiv.org/abs/#1} {{\tt arXiv:#1}}}
\def\mn@eprint@dblp#1{\href {http://dblp.uni-trier.de/rec/bibtex/#1.xml}
  {dblp:#1}}
\def\mn@eprint@#1:#2:#3:#4\@nil{\def\@tempa {#1}\def\@tempb {#2}\def\@tempc
  {#3}\ifx \@tempc \@empty \let \@tempc \@tempb \let \@tempb \@tempa \fi \ifx
  \@tempb \@empty \def\@tempb {arXiv}\fi \@ifundefined
  {mn@eprint@\@tempb}{\@tempb:\@tempc}{\expandafter \expandafter \csname
  mn@eprint@\@tempb\endcsname \expandafter{\@tempc}}}

\bibitem[\protect\citeauthoryear{{Angus}}{{Angus}}{2019}]{2019stardate}
{Angus} R.,  2019, {stardate: a tool for measuring precise stellar ages.},
  Github, \url {https://stardate.readthedocs.io/en/latest/}

\bibitem[\protect\citeauthoryear{Audard, G{\"u}del  \& Skinner}{Audard
  et~al.}{2003}]{audard2003separating}
Audard M.,  G{\"u}del M.,   Skinner S.~L.,  2003, \apj, 589, 983

\bibitem[\protect\citeauthoryear{Barnes}{Barnes}{2007}]{barnes2007ages}
Barnes S.~A.,  2007, \apj, 669, 1167

\bibitem[\protect\citeauthoryear{Barnes, Jeffers, Haswell, Jones, Shulyak,
  Pavlenko  \& Jenkins}{Barnes et~al.}{2017}]{barnes2017surprisingly}
Barnes J.,  Jeffers S.,  Haswell C.,  Jones H.,  Shulyak D.,  Pavlenko Y.~V.,
  Jenkins J.~S.,  2017, \mnras, 471, 811

\bibitem[\protect\citeauthoryear{{Borucki} et~al.,}{{Borucki}
  et~al.}{2010}]{Borucki2010}
{Borucki} W.~J.,  et~al., 2010, \mn@doi [Science] {10.1126/science.1185402},
  \href {http://adsabs.harvard.edu/abs/2010Sci...327..977B} {327, 977}

\bibitem[\protect\citeauthoryear{Byrne, Eibe  \& Van~den Oord}{Byrne
  et~al.}{1998}]{byrne1998evidence}
Byrne P.,  Eibe M.,   Van~den Oord G.,  1998, in International Astronomical
  Union Colloquium. pp 226--234

\bibitem[\protect\citeauthoryear{Davenport et~al.,}{Davenport
  et~al.}{2014}]{davenport2014kepler}
Davenport J.~R.,  et~al., 2014, \apj, 797, 122

\bibitem[\protect\citeauthoryear{Dhillon, Privett  \& Duffey}{Dhillon
  et~al.}{2001}]{dhillon2001period}
Dhillon V.,  Privett G.,   Duffey K.,  2001, Starlink User Note, 167

\bibitem[\protect\citeauthoryear{Doyle, Ramsay, Doyle, Wu  \& Scullion}{Doyle
  et~al.}{2018}]{doyle2018investigating}
Doyle L.,  Ramsay G.,  Doyle J.~G.,  Wu K.,   Scullion E.,  2018, \mnras, 480,
  2153

\bibitem[\protect\citeauthoryear{Eaton, Henry  \& Fekel}{Eaton
  et~al.}{1996}]{eaton1996random}
Eaton J.~A.,  Henry G.~W.,   Fekel F.~C.,  1996, \apj, 462, 888

\bibitem[\protect\citeauthoryear{Fischer \& Saur}{Fischer \&
  Saur}{2019}]{fischer2019time}
Fischer C.,  Saur J.,  2019, \apj, 872, 113

\bibitem[\protect\citeauthoryear{Gaia~Collaboration}{Gaia~Collaboration}{2016}]{gai16}
Gaia~Collaboration Brown A. G. A. e.~a.,  2016, \aap, 595, A2

\bibitem[\protect\citeauthoryear{Gaia~Collaboration}{Gaia~Collaboration}{2018}]{gaia18}
Gaia~Collaboration Brown A. G. A. e.~a.,  2018, \aap, 616, A1

\bibitem[\protect\citeauthoryear{{G{\"u}nther} et~al.,}{{G{\"u}nther}
  et~al.}{2019}]{Gunther2019}
{G{\"u}nther} M.~N.,  et~al., 2019, arXiv e-prints, \href
  {http://adsabs.harvard.edu/abs/2019arXiv190100443G} {}

\bibitem[\protect\citeauthoryear{Hartmann \& Noyes}{Hartmann \&
  Noyes}{1987}]{hartmann1987rotation}
Hartmann L.~W.,  Noyes R.~W.,  1987, Annual review of astronomy and
  astrophysics, 25, 271

\bibitem[\protect\citeauthoryear{Howard, Corbett, Law, Ratzloff, Glazier, Fors,
  del Ser  \& Haislip}{Howard et~al.}{2019}]{howard2019evryflare}
Howard W.~S.,  Corbett H.,  Law N.~M.,  Ratzloff J.~K.,  Glazier A.~L.,  Fors
  O.,  del Ser D.,   Haislip J.,  2019, arXiv preprint arXiv:1904.10421

\bibitem[\protect\citeauthoryear{Kiraga \& Stepien}{Kiraga \&
  Stepien}{2007}]{kiraga2007age}
Kiraga M.,  Stepien K.,  2007, AcA, 57

\bibitem[\protect\citeauthoryear{Kochukhov \& Lavail}{Kochukhov \&
  Lavail}{2017}]{kochukhov2017global}
Kochukhov O.,  Lavail A.,  2017, \apjl, 835, L4

\bibitem[\protect\citeauthoryear{Kuhar, Krucker, Oliveros, Battaglia, Kleint,
  Casadei  \& Hudson}{Kuhar et~al.}{2015}]{kuhar2015correlation}
Kuhar M.,  Krucker S.,  Oliveros J. C.~M.,  Battaglia M.,  Kleint L.,  Casadei
  D.,   Hudson H.~S.,  2015, \apj, 816, 6

\bibitem[\protect\citeauthoryear{Maggio, Sciortino, Vaiana, Majer, Bookbinder,
  Golub, Harnden~Jr  \& Rosner}{Maggio et~al.}{1987}]{maggio1987einstein}
Maggio A.,  Sciortino S.,  Vaiana G.,  Majer P.,  Bookbinder J.,  Golub L.,
  Harnden~Jr F.,   Rosner R.,  1987, \apj, 315, 687

\bibitem[\protect\citeauthoryear{Mariska \& McTiernan}{Mariska \&
  McTiernan}{1999}]{mariska1999hard}
Mariska J.~T.,  McTiernan J.~M.,  1999, \apj, 514, 484

\bibitem[\protect\citeauthoryear{McIntosh}{McIntosh}{1990}]{mcintosh1990classification}
McIntosh P.~S.,  1990, Solar Physics, 125, 251

\bibitem[\protect\citeauthoryear{McQuillan, Aigrain  \& Mazeh}{McQuillan
  et~al.}{2013}]{mcquillan2013measuring}
McQuillan A.,  Aigrain S.,   Mazeh T.,  2013, \mnras, 432, 1203

\bibitem[\protect\citeauthoryear{McQuillan, Mazeh  \& Aigrain}{McQuillan
  et~al.}{2014}]{mcquillan2014rotation}
McQuillan A.,  Mazeh T.,   Aigrain S.,  2014, \apjs, 211, 24

\bibitem[\protect\citeauthoryear{{Morton}}{{Morton}}{2015}]{2015isochrones}
{Morton} T.~D.,  2015, {isochrones: Stellar model grid package}, Astrophysics
  Source Code Library (\mn@eprint {ascl} {1503.010})

\bibitem[\protect\citeauthoryear{Ol{\'a}h, Kov{\'a}ri, Bartus, Strassmeier,
  Hall  \& Henry}{Ol{\'a}h et~al.}{1997}]{olah1997time}
Ol{\'a}h K.,  Kov{\'a}ri Z.,  Bartus J.,  Strassmeier K.,  Hall D.,   Henry G.,
   1997, \aap, 321, 811

\bibitem[\protect\citeauthoryear{{Ramsay}, {Doyle}, {Hakala}, {Garcia-Alvarez},
  {Brooks}, {Barclay}  \& {Still}}{{Ramsay} et~al.}{2013}]{ramsay2013short}
{Ramsay} G.,  {Doyle} J.~G.,  {Hakala} P.,  {Garcia-Alvarez} D.,  {Brooks} A.,
  {Barclay} T.,   {Still} M.,  2013, \mn@doi [\mnras] {10.1093/mnras/stt1182},
  \href {http://adsabs.harvard.edu/abs/2013MNRAS.434.2451R} {434, 2451}

\bibitem[\protect\citeauthoryear{{Ricker} et~al.,}{{Ricker}
  et~al.}{2015}]{Ricker2015}
{Ricker} G.~R.,  et~al., 2015, \mn@doi [Journal of Astronomical Telescopes,
  Instruments, and Systems] {10.1117/1.JATIS.1.1.014003}, \href
  {http://adsabs.harvard.edu/abs/2015JATIS...1a4003R} {1, 014003}

\bibitem[\protect\citeauthoryear{Roettenbacher \& Vida}{Roettenbacher \&
  Vida}{2018}]{roettenbacher2018connection}
Roettenbacher R.~M.,  Vida K.,  2018, \apj, 868, 3

\bibitem[\protect\citeauthoryear{Route}{Route}{2019}]{route2019rise}
Route M.,  2019, \apj, 872, 79

\bibitem[\protect\citeauthoryear{{Silva}}{{Silva}}{2003}]{silva2003}
{Silva} A.~V.~R.,  2003, \mn@doi [\apjl] {10.1086/374324}, \href
  {https://ui.adsabs.harvard.edu/abs/2003ApJ...585L.147S} {585, L147}

\bibitem[\protect\citeauthoryear{Stassun et~al.,}{Stassun
  et~al.}{2018}]{stassun2018tess}
Stassun K.~G.,  et~al., 2018, The Astronomical Journal, 156, 102

\bibitem[\protect\citeauthoryear{Strassmeier}{Strassmeier}{1996}]{strassmeier1996observational}
Strassmeier K.~G.,  1996, in Symposium-International Astronomical Union. pp
  289--298

\bibitem[\protect\citeauthoryear{Van~den Oord, Byrne  \& Eibe}{Van~den Oord
  et~al.}{1998}]{van1998prominence}
Van~den Oord G.,  Byrne P.,   Eibe M.,  1998, in International Astronomical
  Union Colloquium. pp 251--254

\bibitem[\protect\citeauthoryear{Wolf et~al.,}{Wolf
  et~al.}{2018}]{wolf2018skymapper}
Wolf C.,  et~al., 2018, Publications of the Astronomical Society of Australia,
  35

\bibitem[\protect\citeauthoryear{Yang et~al.,}{Yang
  et~al.}{2017}]{yang2017flaring}
Yang H.,  et~al., 2017, \apj, 849, 36

\bibitem[\protect\citeauthoryear{Zirin \& Liggett}{Zirin \&
  Liggett}{1982}]{zirin1982delta}
Zirin H.,  Liggett M.~A.,  1982, Solar Physics, 113, 267

\makeatother
\end{thebibliography}

\appendix

\section{}
\begin{table*}
\caption{The stellar properties of a select few stars which showed no rotational modulation but processed flares within their lightcurves, detailing the number of flares, quiescent luminosity, energy range and duration range of the flares. The apparent magnitude in the TESS band-pass, $T_{mag}$, is taken from the TESS Input Catalog (TIC) along with the TIC ID \citep{stassun2018tess}. The distances are derived from the Gaia Data Release 2 parallaxes \citep{gai16, gaia18} and the spectral types are obtained from the {\tt SIMBAD} catalogue. \newline
{\it This table is available in its entirety in a machine-readable form in the online journal. A portion is shown here for guidance regarding its form and content.}}

   \begin{center}
   \label{no_mod_flares}
\resizebox{\textwidth}{!}{
	\begin{tabular}{lccccccccccc}
    \hline 
	Name                  & TIC ID     &  Sector     & No. of   & SpT    & $T_{mag}$  &  Parallax             &  Distance             &  $log(L_{star})$  & $log(E_{flare})$   &  Duration        \\
	                      &            &             & Flares   &        &            &  mas                  &  pc                   &  erg/s            & erg             &  minutes         \\
	\hline
	2MASS J2148-4736      & 147421845  &  1          & 3        &  5.0   & 12.694     & $17.869 \pm 0.064$   & $55.9623 \pm 0.2023$  &  31.18            & 33.00 -- 33.53  & 42.00 -- 47.99    \\
    LP873-37              & 099566892  &  1          & 5        &  4.0   & 12.12      & $21.106 \pm 0.076$   & $47.3165 \pm 0.1706$  &  31.26            & 32.59 -- 33.37  & 17.99 -- 80.00    \\
    2XMM J2253-1721       & 188586529  &  2          & 10       &  4.0   & 12.948     & $24.485 \pm 0.124$   & $40.8420 \pm 0.2068$  &  30.82            & 32.29 -- 33.36  & 17.99 -- 63.99    \\
    WISE J0127-6032       & 237910557  &  1\&2       & 7        &  4.2   & 12.824     & $19.770 \pm 0.058$   & $50.5845 \pm 0.1501$  &  31.05            & 31.91 -- 32.90  & 12.00 -- 54.00    \\
    2MASS J0413-5231      & 219229275  &  3          & 2        &  2.4   & 11.54      & $19.428 \pm 0.036$   & $51.4721 \pm 0.0967$  &  31.58            & 33.54 -- 33.63  & 104.00 -- 128.00   \\
    PS 78191              & 011652986  &  3          & 5        &  3.5   & 11.58      & $29.532 \pm 0.074$   & $33.8608 \pm 0.0857$  &  31.19            & 32.21 -- 33.20  & 8.00 -- 37.99  \\
    \hline
    \end{tabular}}
    \end{center}
\end{table*}

% Don't change these lines
\bsp	% typesetting comment
\label{lastpage}
\end{document}